\documentclass[12pt]{article}\usepackage{latexsym}%
\newlength\Cscr\newlength\Csave
\newlength\Ctenthex\newlength\CFxsize
\newlength\CFxsizeps\newlength\CFsizemakebox\newlength\CFleftcrop
\newlength\CFrightcrop\newlength\CZtbldist\newlength\CZfigdist
\newlength
\CGDnum\newlength\CGDtext\newcounter{Cscr}%
\newcounter{CBcit}%
\newcounter{CBauthornu%
m}\newcounter{Ceqindent}\newcounter{CBtnc}
\newcounter{CBtntc}\newcounter{CEht}%
\newcounter{Cbscurr}\newcounter{CbsA}\newcounter{CbsB}\newcounter
{CbsC}\newcounter{CbsD}\hoffset\Cscr\voffset
\Cscr\protect
\begin{document}\setcounter{figure}{1}\setcounter{t%
able}{0}\renewcommand\theequation{\arabic{equation}}\renewcommand
\thetable{\arabic{table}}\renewcommand\thefigure{\arabic{figure}%
}\renewcommand\thesection{\arabic{section}}\renewcommand
\thesubsection{\arabic{section}.\arabic{subsection}}\renewcommand
\thesubsubsection{\arabic{section}.\arabic{subsection}.\arabic{s%
ubsubsection}}\setcounter{CEht}{10}\setcounter{CbsA}{1}%
\setcounter{CbsB}{1}\setcounter{CbsC}{1}\setcounter{CbsD}{1}{%
\centering{\protect\mbox{}}\\*[\baselineskip]{\large\bf A new la%
ttice gauge action without scaling artifacts}\\*}{\centering{%
\protect\mbox{}}\\John P.~Costella\vspace{1ex}\\*}{\centering{%
\small\protect\emph{Narre Warren, Melbourne, Australia}}\\}{%
\centering{\small\protect\emph{jpcostella@hotmail.com; www.users%
.bigpond.com/costella}}\\}{\centering{\protect\mbox{}}\\(October%
\ 3, 2004\vspace{1ex})\\}\par\vspace\baselineskip\setlength{%
\Csave}{\parskip}{\centering\small\bf Abstract\\}\setlength{%
\parskip}{-\baselineskip}\begin{quote}\setlength{\parskip}{0.5%
\baselineskip}\small\noindent I describe a new way of constructi%
ng a gauge action that eliminates scaling artifacts, by writing 
the continuum formalism in terms of ``gauge links'' (Schwinger l%
ine integrals) and using the optimal SLAC representation of the 
lattice derivative. Computational performance can be maintained 
by implementing the action as a ``stochastic'' operator, as has 
been recently implemented in the MILC code for the optimal $D$-s%
lash operator and the antialiasing filter. \end{quote}\setlength
{\parskip}{\Csave}\par\refstepcounter{section}\vspace{1.5%
\baselineskip}\par{\centering\bf\thesection. Introduction\\*[0.5%
\baselineskip]}\protect\indent\label{sect:Introduction}I recentl%
y proposed a new paradigm for lattice calculations that addresse%
s a number of fundamental flaws in current practice. Until now, 
fields have been multiplied together in position space without a%
ny pre-filtering or post-filtering. But it is a fundamental resu%
lt of Fourier theory that such unfiltered products will automati%
cally ``foul themselves''\mbox{$\!$}, because high-momentum mode%
s in the product ``fold back'' from the Brillouin zone boundary 
and ``masquerade'' or ``alias'' as low-momentum modes (see, 
\protect\emph{e.g.}, [\ref{cit:Press1988}]). This ``fouling'' oc%
curs every time that a product of fields is formed---which in th%
e context of lattice field theory means at practically every ste%
p. That practitioners have managed to extract meaningful numbers 
from the resulting mess is close to miraculous---and it has only 
been achieved by using enormously large lattices and intricate e%
xtrapolation procedures.\par Some have discovered, empirically, 
that high momentum modes constitute more noise than signal, beca%
use the more they ``fatten'' or ``smear'' the links and operator%
s, the better the signal-to-noise ratio has become. But the corr%
ect solution, that eliminates the mess before it happens, is to 
apply an ``antialiasing filter'' to each field, at each and ever%
y stage, that has no effect on any momentum mode with \raisebox{%
0ex}[1ex][0ex]{$\protect\displaystyle|p_\mu|<2\pi/3a$}, but elim%
inates all modes outside this window~[\ref{cit:Costella2004}]. S%
uch a filter has now been implemented, for both scalar and paral%
lel computation, using the public-domain MILC code~[\ref{cit:MIL%
C}], with performance being maintained by implementing the opera%
tor in a ``stochastic'' fashion~[\ref{cit:Costella2002a}], where 
we can ``dial'' the stochasticity from zero (fully deterministic%
) to as high as we like (every coefficient interpreted as a prob%
ability).\par Likewise, until now the representation of fermions 
on the lattice has been woefully inadequate. The use of ``ultral%
ocal'' derivative operators was originally mandated by performan%
ce considerations, and by the Nielson--Ninomiya theorem locked p%
ractitioners into a no-win situation of having to deal with eith%
er doubled fermions or destroyed chiral properties. The developm%
ent of Ginsparg--Wilson fermions has begun to break down the myt%
h of the \protect\emph{need} for ultralocality. However, I have 
argued~[\ref{cit:Costella2002a},\,\ref{cit:Costella2002b}]{} tha%
t the correct way to proceed is to simply use the ``SLAC'' deriv%
ative operator~[\ref{cit:Drell1976}], which is defined to have a 
perfect representation in momentum space (within the first Brill%
ouin zone). The SLAC operator seems to have been tried and disca%
rded long ago, and there is resistance to it being resurrected. 
However, the reason for it being discredited is now easy to disc%
ern. By definition, the momentum modes of the SLAC derivative op%
erator with largest magnitude are those near the Brillouin zone 
boundary (simply because it is defined to be $ip_\mu$ in momentu%
m space). Without any antialiasing filters, these dominant modes 
will be aliased right down to low momentum values, with just a s%
ingle $D$-slash operation. Extracting any meaningful signal from 
beneath this aliasing noise becomes impossible. Ironically, the 
very weakness of an ultralocal derivative operator that causes t%
he doubling problem in the first place---the fact that it must v%
anish \protect\emph{somewhere} other than \raisebox{0ex}[1ex][0e%
x]{$\protect\displaystyle p=0$}, usually at the Brillouin zone b%
oundary---coincidentally provides an (incomplete) filtering of h%
igh-momentum modes, which makes its performance appear superior 
to the SLAC operator, if no antialiasing filters are applied.\par
Removing this illusory disadvantage with a properly antialiased 
lattice engine, we still have to face the question of performanc%
e and cost. The SLAC derivative operator connects every site to 
every other site along the direction of the derivative. Here, ag%
ain, the implementation of the operator in a ``stochastic'' fash%
ion solves the problem, allowing the operator to perform perfect%
ly in the mean, where we can dial the amount of ``stochasticity%
'' by trading performance off against noise, ultimately finding 
the setting that optimizes our computing power. This optimal $D$%
-slash operator has also been recently implemented for both scal%
ar and parallel computations, using the MILC code~[\ref{cit:MILC%
}].\par The construction of these two important planks in the ne%
w lattice paradigm begs a tantalizing question: Is it possible t%
o construct a lattice engine that has \protect\emph{no} inherent 
scaling artifacts at all? Such a computation engine would be the 
Holy Grail of lattice research. The only effects of the finitene%
ss of the lattice would be physical ones: the ``squeezing'' of s%
tates (if the lattice is too small, in its overall dimensions), 
and the elimination (regulation) of high-momentum modes, for any 
finite value of~$a$. The former is expected to fall off exponent%
ially, provided that the overall dimensions of the lattice are s%
ufficiently large (a few fm or more in each direction); the latt%
er represents the effects of regulation and renormalization, tha%
t are fundamental to the physical description.\par Can such an e%
ngine be constructed? With an optimal $D$-slash operator, and an 
antialiasing filter, the only part of a lattice computation reta%
ining scaling artifacts is the evaluation of the gauge action. N%
ow, the lowest-approximation plaquette estimate has been ``impro%
ved'' in numerous ways over the years, but, again, the gospel of 
ultralocality has dominated. Generally, a few higher-order strin%
gs of links have been introduced, with their coefficients manual%
ly matched up to provide a cancellation of artifacts to some ord%
er in~$a$. Some of these improvements also mix in the dynamics, 
to provide better cancellation.\par But is there no simple way t%
o systematically \protect\emph{derive} all of the required strin%
gs of links? Consider: if we actually represented the gauge fiel%
d by the $A_\mu$, rather than links, then all we would need to d%
o would be to compute \raisebox{0ex}[1ex][0ex]{$\protect
\displaystyle F_{\!\mu\nu}$} by means of derivative operations (%
and a commutator), and we already know the optimal way to perfor%
m a derivative (the SLAC operator). Moreover, this would be a ma%
nifestly two-dimensional problem: you only compute this ``curl''%
\mbox{$\!$}, namely, \raisebox{0ex}[1ex][0ex]{$\protect
\displaystyle F_{\!\mu\nu}$}, two dimensions at a time. Of cours%
e, we \protect\emph{don't} represent the gauge field by the $A_%
\mu$, because we want to maintain gauge invariance identically%
---that's why we use the links. But the links are derived fairly 
simply from the $A_\mu$; so is there a way to algorithmically co%
nstruct a gauge action with a perfect dispersion relation (withi%
n the first Brillouin zone) by using derivative operations on th%
e link description itself?\par In the following sections I will 
argue that the answer to this question is ``yes''\mbox{$\!$}. In 
\mbox{Sec.~$\:\!\!$}\protect\ref{sect:Continuum} I show how the 
\protect\emph{continuum} gauge theory can be rewritten in terms 
of ``links'' (\protect\emph{i.e.}, Schwinger line integrals) ins%
tead of the $A_\mu$, where derivative operations are applied to 
shift the endpoints of these links infinitesimally. In \mbox{Sec%
.~$\:\!\!$}\protect\ref{sect:Lattice} I then show how the SLAC d%
erivative operator can be used to optimally represent this conti%
nuum description on the lattice, with each contribution being a 
closed string of gauge links; in other words, gauge invariance i%
s maintained for each and every contribution. The evaluation of 
these closed strings can then be performed stochastically, in th%
e same way that the antialiasing and optimal $D$-slash operation%
s have already been implemented in the MILC code.\par
\refstepcounter{section}\vspace{1.5\baselineskip}\par{\centering
\bf\thesection. Continuum gauge theory using links\\*[0.5%
\baselineskip]}\protect\indent\label{sect:Continuum}Consider the 
Schwinger line integral \raisebox{0ex}[1ex][0ex]{$\protect
\displaystyle U^{\:\!\!}(y,x,^{\!}P)$} along a path $P$ between 
spacetime points $x$ and $y$: \setcounter{Ceqindent}{0}\protect
\begin{eqnarray}\protect\left.\protect\begin{array}{rcl}\protect
\displaystyle\hspace{-1.3ex}&\protect\displaystyle U^{\:\!\!}(y,%
x,^{\!}P)\equiv{\cal P}\exp\!\setcounter{Cbscurr}{25}\setlength{%
\Cscr}{\value{Cbscurr}\Ctenthex}\addtolength{\Cscr}{-1.0ex}%
\protect\raisebox{0ex}[\value{Cbscurr}\Ctenthex][\Cscr]{}\hspace
{-0ex}{\protect\left(\setlength{\Cscr}{\value{Cbscurr}\Ctenthex}%
\addtolength{\Cscr}{-1.0ex}\protect\raisebox{0ex}[\value{Cbscurr%
}\Ctenthex][\Cscr]{}\protect\right.}\hspace{-0.25ex}\setlength{%
\Cscr}{\value{Cbscurr}\Ctenthex}\addtolength{\Cscr}{-1.0ex}%
\protect\raisebox{0ex}[\value{Cbscurr}\Ctenthex][\Cscr]{}%
\setcounter{CbsD}{\value{CbsC}}\setcounter{CbsC}{\value{CbsB}}%
\setcounter{CbsB}{\value{CbsA}}\setcounter{CbsA}{\value{Cbscurr}%
}ig\hspace{-0.5mm}{\protect\mbox{}}\hspace{-0.1mm}\makebox[0ex]{%
\raisebox{-5.4mm}{\hspace{-0ex}\hspace{1mm}\makebox[0ex]{%
\scriptsize{\protect\mbox{}}$P$}}\hspace{0ex}\hspace{-1.5mm}}%
\protect\int_{x}^{y}{\protect\mbox{}}\hspace{-0.5mm}\hspace{-0.6%
mm}dz^\mu A_{\mu\:\!\!}(z)\setlength{\Cscr}{\value{CbsA}\Ctenthex
}\addtolength{\Cscr}{-1.0ex}\protect\raisebox{0ex}[\value{CbsA}%
\Ctenthex][\Cscr]{}\hspace{-0.25ex}{\protect\left.\setlength{%
\Cscr}{\value{CbsA}\Ctenthex}\addtolength{\Cscr}{-1.0ex}\protect
\raisebox{0ex}[\value{CbsA}\Ctenthex][\Cscr]{}\protect\right)}%
\hspace{-0ex}\setlength{\Cscr}{\value{CbsA}\Ctenthex}\addtolength
{\Cscr}{-1.0ex}\protect\raisebox{0ex}[\value{CbsA}\Ctenthex][%
\Cscr]{}\setcounter{CbsA}{\value{CbsB}}\setcounter{CbsB}{\value{%
CbsC}}\setcounter{CbsC}{\value{CbsD}}\setcounter{CbsD}{1},%
\setlength{\Cscr}{\value{CEht}\Ctenthex}\addtolength{\Cscr}{-1.0%
ex}\protect\raisebox{0ex}[\value{CEht}\Ctenthex][\Cscr]{}\protect
\end{array}\protect\right.\protect\label{eq:Continuum-Schwinger}%
\protect\end{eqnarray}\setcounter{CEht}{10}where \setcounter{Ceq%
indent}{0}\protect\begin{eqnarray}\hspace{-1.3ex}&\displaystyle
A_{\mu\:\!\!}(x)\equiv T_{b\:\!}^{\protect\mbox{}}A^b_{\mu\:\!\!%
}(x)\protect\nonumber\setlength{\Cscr}{\value{CEht}\Ctenthex}%
\addtolength{\Cscr}{-1.0ex}\protect\raisebox{0ex}[\value{CEht}%
\Ctenthex][\Cscr]{}\protect\end{eqnarray}\setcounter{CEht}{10}is 
the gauge field, the \raisebox{0ex}[1ex][0ex]{$\protect
\displaystyle T_b^{\protect\mbox{}}$} are the generators of the 
gauge group, $g$ is the bare coupling constant of the gauge inte%
raction, and the path-ordering operator ${\cal P}$ ensures that 
products of non-Abelian quantities arising from the exponentiati%
on operation are correctly arranged according to the order that 
they are encountered along the path of integration. The links of 
lattice gauge theory are simply Schwinger line integrals of leng%
th $a$ along the four principal lattice directions; in other wor%
ds, Schwinger line integrals are ``continuum links''\mbox{$\!$}.%
\par Under a gauge transformation, \raisebox{0ex}[1ex][0ex]{$%
\protect\displaystyle U^{\:\!\!}(y,x,^{\!}P)$} transforms at the 
endpoints $x$ and $y$, oppositely to $\psi(x)$ and $\bar\psi(y)$ 
respectively, so that the quantity \setcounter{Ceqindent}{0}%
\protect\begin{eqnarray}\protect\left.\protect\begin{array}{rcl}%
\protect\displaystyle\hspace{-1.3ex}&\protect\displaystyle\bar
\psi(y)^{\:\!}U^{\:\!\!}(y,x,^{\!}P)^{\:\!}\psi(x)\setlength{%
\Cscr}{\value{CEht}\Ctenthex}\addtolength{\Cscr}{-1.0ex}\protect
\raisebox{0ex}[\value{CEht}\Ctenthex][\Cscr]{}\protect\end{array%
}\protect\right.\protect\label{eq:Continuum-CappedLink}\protect
\end{eqnarray}\setcounter{CEht}{10}is gauge-invariant, as is 
\setcounter{Ceqindent}{0}\protect\begin{eqnarray}\protect\left.%
\protect\begin{array}{rcl}\protect\displaystyle\hspace{-1.3ex}&%
\protect\displaystyle\mbox{Tr$\:$}U^{\:\!\!}(x,x,^{\!}P),%
\setlength{\Cscr}{\value{CEht}\Ctenthex}\addtolength{\Cscr}{-1.0%
ex}\protect\raisebox{0ex}[\value{CEht}\Ctenthex][\Cscr]{}\protect
\end{array}\protect\right.\protect\label{eq:Continuum-LoopedLink%
}\protect\end{eqnarray}\setcounter{CEht}{10}for any closed path 
$P$ leading from $x$ back to $x$.\par Consider, now, the ``parti%
al derivative'' \setcounter{Ceqindent}{0}\protect\begin{eqnarray%
}\hspace{-1.3ex}&\displaystyle\mbox{$\protect\displaystyle
\protect\frac{\partial}{\partial x^\mu}$}^{\:\!}U^{\:\!\!}(y,x,^%
{\!}P),\protect\nonumber\setlength{\Cscr}{\value{CEht}\Ctenthex}%
\addtolength{\Cscr}{-1.0ex}\protect\raisebox{0ex}[\value{CEht}%
\Ctenthex][\Cscr]{}\protect\end{eqnarray}\setcounter{CEht}{10}ob%
tained by shifting the endpoint $x$ by a small amount \raisebox{%
0ex}[1ex][0ex]{$\protect\displaystyle\delta x^{\mu\!}$} in the $%
\mu$-direction, with the $x$-end of the path $P$ being extended 
by this additional displacement \raisebox{0ex}[1ex][0ex]{$%
\protect\displaystyle\delta^{\:\!\!}P^{\mu\!}\equiv\hat\mu\,%
\delta x^\mu$}\mbox{$\!$}, and using first principles to compute 
the ``derivative'': \setcounter{Ceqindent}{0}\protect\begin{eqna%
rray}\protect\left.\protect\begin{array}{rcl}\protect
\displaystyle\hspace{-1.3ex}&\protect\displaystyle\mbox{$\protect
\displaystyle\protect\frac{\partial}{\partial x^\mu}$}^{\:\!}U^{%
\:\!\!}(y,x,^{\!}P)\equiv\lim_{\delta x^{\mu\!}\rightarrow0}\!%
\mbox{$\protect\displaystyle\protect\frac{U^{\:\!\!}(y,^{\:\!}x%
\!+\!\delta x^{\mu\:\!\!},^{\:\!}P\:\!\!\!+\!\delta^{\:\!\!}P^{%
\mu\!})-U^{\:\!\!}(y,x,^{\!}P)}{\delta x^{\mu\!}}$}.\setlength{%
\Cscr}{\value{CEht}\Ctenthex}\addtolength{\Cscr}{-1.0ex}\protect
\raisebox{0ex}[\value{CEht}\Ctenthex][\Cscr]{}\protect\end{array%
}\protect\right.\protect\label{eq:Continuum-PathPartialDerivativ%
e}\protect\end{eqnarray}\setcounter{CEht}{10}It is clear from (%
\protect\ref{eq:Continuum-Schwinger}) that \setcounter{Ceqindent%
}{0}\protect\begin{eqnarray}\protect\left.\protect\begin{array}{%
rcl}\protect\displaystyle\hspace{-1.3ex}&\protect\displaystyle
\mbox{$\protect\displaystyle\protect\frac{\partial}{\partial x^%
\mu}$}^{\:\!}U^{\:\!\!}(y,x,^{\!}P)=-ig^{\,}U^{\:\!\!}(y,x,^{\!}%
P)^{\,}A_{\mu\:\!\!}(x),\setlength{\Cscr}{\value{CEht}\Ctenthex}%
\addtolength{\Cscr}{-1.0ex}\protect\raisebox{0ex}[\value{CEht}%
\Ctenthex][\Cscr]{}\protect\end{array}\protect\right.\protect
\label{eq:Continuum-dUdx}\protect\end{eqnarray}\setcounter{CEht}%
{10}from the fundamental definition of integration, and the path%
-ordering operator ${\cal P}$.\par Consider, now, the quantity 
\setcounter{Ceqindent}{0}\protect\begin{eqnarray}\protect\left.%
\protect\begin{array}{rcl}\protect\displaystyle\hspace{-1.3ex}&%
\protect\displaystyle C_{\:\!\!F}(x)\equiv i\mbox{$\protect
\displaystyle\protect\frac{\partial}{\partial x^\mu}$}^{\:\!}\bar
\psi(y)^{\:\!}\gamma^{\mu\,}U^{\:\!\!}(y,x,^{\!}P)^{\:\!}\psi(x)%
\setlength{\Cscr}{\value{Cbscurr}\Ctenthex}\addtolength{\Cscr}{-%
1.0ex}\protect\raisebox{0ex}[\value{Cbscurr}\Ctenthex][\Cscr]{}%
\hspace{-0.05ex}{\protect\left|\setlength{\Cscr}{\value{Cbscurr}%
\Ctenthex}\addtolength{\Cscr}{-1.0ex}\protect\raisebox{0ex}[%
\value{Cbscurr}\Ctenthex][\Cscr]{}\protect\right.}\hspace{-0.25e%
x}\setlength{\Cscr}{\value{Cbscurr}\Ctenthex}\addtolength{\Cscr}%
{-1.0ex}\protect\raisebox{0ex}[\value{Cbscurr}\Ctenthex][\Cscr]{%
}_{y\rightarrow x,^{\,}P^{\:\!\!}\rightarrow0\displaystyle,}%
\setlength{\Cscr}{\value{CEht}\Ctenthex}\addtolength{\Cscr}{-1.0%
ex}\protect\raisebox{0ex}[\value{CEht}\Ctenthex][\Cscr]{}\protect
\end{array}\protect\right.\protect\label{eq:Continuum-DefineCFx}%
\protect\end{eqnarray}\setcounter{CEht}{10}where the point $y$ i%
s brought into coincidence with the point $x$ after the differen%
tiation with respect to \raisebox{0ex}[1ex][0ex]{$\protect
\displaystyle x^{\mu\!}$} has been performed, and where in writi%
ng \raisebox{0ex}[1ex][0ex]{$\protect\displaystyle P^{\:\!\!}%
\rightarrow0$} we are implying that the length of the path $P$ f%
rom $x$ to $y$ shrinks to zero as $y$ approaches $x$; \protect
\emph{i.e.}, there is no ``loop'' of $P$ left outside the point 
$x$ when we take the limit. Using the product rule of differenti%
ation, and noting that \raisebox{0ex}[1ex][0ex]{$\protect
\displaystyle\bar\psi(y)$} is independent of $x$, we find that (%
\protect\ref{eq:Continuum-DefineCFx}) becomes \setcounter{Ceqind%
ent}{0}\protect\begin{eqnarray}\hspace{-1.3ex}&\displaystyle C_{%
\:\!\!F}(x)=i\!\protect\left\{\bar\psi(y)^{\,}\gamma^{\mu\:\!\!}%
\!\protect\left[\mbox{$\protect\displaystyle\protect\frac{%
\partial}{\partial x^\mu}$}^{\:\!}U^{\:\!\!}(y,x,^{\!}P)\protect
\right]\!\psi(x)+\bar\psi(y)^{\:\!}\gamma^{\mu\,}U^{\:\!\!}(y,x,%
^{\!}P)\mbox{$\protect\displaystyle\protect\frac{\partial}{%
\partial x^\mu}$}^{\:\!}\psi(x)\protect\right\}_{\!y\rightarrow
x,^{\,}P^{\:\!\!}\rightarrow0\displaystyle.}\protect\nonumber
\setlength{\Cscr}{\value{CEht}\Ctenthex}\addtolength{\Cscr}{-1.0%
ex}\protect\raisebox{0ex}[\value{CEht}\Ctenthex][\Cscr]{}\protect
\end{eqnarray}\setcounter{CEht}{10}Substituting the result (%
\protect\ref{eq:Continuum-dUdx}) for the expression in brackets, 
we then find that \setcounter{Ceqindent}{0}\protect\begin{eqnarr%
ay}\hspace{-1.3ex}&\displaystyle C_{\:\!\!F}(x)=\protect\left\{g%
^{\,}\bar\psi(y)^{\,}\gamma^{\mu\,}U(y,x,^{\!}P)^{\,}A_\mu(x)^{%
\,}\psi(x)+i^{\:\!}\bar\psi(y)^{\:\!}\gamma^{\mu\,}U^{\:\!\!}(y,%
x,^{\!}P)\mbox{$\protect\displaystyle\protect\frac{\partial}{%
\partial x^\mu}$}^{\:\!}\psi(x)\protect\right\}_{\!y\rightarrow
x,^{\,}P^{\:\!\!}\rightarrow0\displaystyle.}\protect\nonumber
\setlength{\Cscr}{\value{CEht}\Ctenthex}\addtolength{\Cscr}{-1.0%
ex}\protect\raisebox{0ex}[\value{CEht}\Ctenthex][\Cscr]{}\protect
\end{eqnarray}\setcounter{CEht}{10}The quantity \raisebox{0ex}[1%
ex][0ex]{$\protect\displaystyle U^{\:\!\!}(y,x,^{\!}P)$} is now 
no longer being differentiated, and so we can immediately take t%
he limit \raisebox{0ex}[1ex][0ex]{$\protect\displaystyle y%
\rightarrow x$} with \raisebox{0ex}[1ex][0ex]{$\protect
\displaystyle P^{\:\!\!}\rightarrow0$}. Noting that \setcounter{%
Ceqindent}{0}\protect\begin{eqnarray}\hspace{-1.3ex}&%
\displaystyle U^{\:\!\!}(y,x,^{\!}P)\setlength{\Cscr}{\value{Cbs%
curr}\Ctenthex}\addtolength{\Cscr}{-1.0ex}\protect\raisebox{0ex}%
[\value{Cbscurr}\Ctenthex][\Cscr]{}\hspace{-0.05ex}{\protect\left
|\setlength{\Cscr}{\value{Cbscurr}\Ctenthex}\addtolength{\Cscr}{%
-1.0ex}\protect\raisebox{0ex}[\value{Cbscurr}\Ctenthex][\Cscr]{}%
\protect\right.}\hspace{-0.25ex}\setlength{\Cscr}{\value{Cbscurr%
}\Ctenthex}\addtolength{\Cscr}{-1.0ex}\protect\raisebox{0ex}[%
\value{Cbscurr}\Ctenthex][\Cscr]{}_{y\rightarrow x,^{\,}P^{\:\!%
\!}\rightarrow0\!}\equiv1,\protect\nonumber\setlength{\Cscr}{%
\value{CEht}\Ctenthex}\addtolength{\Cscr}{-1.0ex}\protect
\raisebox{0ex}[\value{CEht}\Ctenthex][\Cscr]{}\protect\end{eqnar%
ray}\setcounter{CEht}{10}we therefore find that \setcounter{Ceqi%
ndent}{0}\protect\begin{eqnarray}\protect\left.\protect\begin{ar%
ray}{rcl}\protect\displaystyle\hspace{-1.3ex}&\protect
\displaystyle C_{\:\!\!F}(x)=i^{\:\!}\bar\psi(x)(\mbox{$\partial
\hspace{-1.15ex}\hspace{0.0ex}\raisebox{0.2ex}\slash\hspace{-0.0%
ex}\hspace{0.03ex}$}\!-\!ig\mbox{$A\hspace{-1.15ex}\hspace{-0.1e%
x}\raisebox{0.2ex}\slash\hspace{--0.1ex}\hspace{0.03ex}$})\psi(x%
)\equiv i^{\:\!}\bar\psi(x)\mbox{$D\hspace{-1.15ex}\hspace{-0.25%
ex}\raisebox{0.2ex}\slash\hspace{--0.25ex}\hspace{0.03ex}$}^{\:%
\!}\psi(x),\setlength{\Cscr}{\value{CEht}\Ctenthex}\addtolength{%
\Cscr}{-1.0ex}\protect\raisebox{0ex}[\value{CEht}\Ctenthex][\Cscr
]{}\protect\end{array}\protect\right.\protect\label{eq:Continuum%
-CFxDslash}\protect\end{eqnarray}\setcounter{CEht}{10}namely, th%
e fermion $D$-slash part of the Lagrangian.\par\mbox{Eq.~%
\raisebox{0ex}[1ex][0ex]{$\protect\displaystyle\!$}}(\protect\ref
{eq:Continuum-DefineCFx}) thus tells us how to construct the fer%
mion part of the continuum Lagrangian out of links, rather than 
the $A_\mu$: Join $\bar\psi(y)$ to $\psi(x)$ with a $\gamma^\mu$ 
and a gauge link from $y$ to $x$ sandwiched in between, then tak%
e the ``partial derivative'' \raisebox{0ex}[1ex][0ex]{$\protect
\displaystyle\partial/\partial x^{\mu\!}$}, \protect\emph{with t%
he link attached}, and then finally bring $y$ into coincidence w%
ith $x$.\par Now consider \setcounter{Ceqindent}{0}\protect\begin
{eqnarray}\protect\left.\protect\begin{array}{rcl}\protect
\displaystyle\hspace{-1.3ex}&\protect\displaystyle\hspace{-5ex}C%
_{\:\!\!G}(x)\equiv\protect\left(\!\mbox{$\protect\displaystyle
\protect\frac{\partial}{\partial x^\mu}$}\mbox{$\protect
\displaystyle\protect\frac{\partial}{\partial x^\nu}$}\!-\!\mbox
{$\protect\displaystyle\protect\frac{\partial}{\partial x^\nu}$}%
\mbox{$\protect\displaystyle\protect\frac{\partial}{\partial x^%
\mu}$}\!^{\:\!\!}\protect\right){}^{\:\!\!}\!\!\!\protect\left(%
\!\mbox{$\protect\displaystyle\protect\frac{\partial}{\partial y%
_{\:\!\!\mu}}$}\mbox{$\protect\displaystyle\protect\frac{\partial
}{\partial y_{\:\!\!\nu}}$}\!-\!\mbox{$\protect\displaystyle
\protect\frac{\partial}{\partial y_{\:\!\!\nu}}$}\mbox{$\protect
\displaystyle\protect\frac{\partial}{\partial y_{\:\!\!\mu}}$}\!%
^{\:\!\!}\protect\right)\!\mbox{Tr$\:$}U^{\:\!\!}(y,x,^{\!}P)%
\setlength{\Cscr}{\value{Cbscurr}\Ctenthex}\addtolength{\Cscr}{-%
1.0ex}\protect\raisebox{0ex}[\value{Cbscurr}\Ctenthex][\Cscr]{}%
\hspace{-0.05ex}{\protect\left|\setlength{\Cscr}{\value{Cbscurr}%
\Ctenthex}\addtolength{\Cscr}{-1.0ex}\protect\raisebox{0ex}[%
\value{Cbscurr}\Ctenthex][\Cscr]{}\protect\right.}\hspace{-0.25e%
x}\setlength{\Cscr}{\value{Cbscurr}\Ctenthex}\addtolength{\Cscr}%
{-1.0ex}\protect\raisebox{0ex}[\value{Cbscurr}\Ctenthex][\Cscr]{%
}_{y\rightarrow x,^{\,}P^{\:\!\!}\rightarrow0\displaystyle,}%
\hspace{-5ex}\setlength{\Cscr}{\value{CEht}\Ctenthex}\addtolength
{\Cscr}{-1.0ex}\protect\raisebox{0ex}[\value{CEht}\Ctenthex][%
\Cscr]{}\protect\end{array}\protect\right.\protect\label{eq:Cont%
inuum-DefineCGx}\protect\end{eqnarray}\setcounter{CEht}{10}which 
is gauge-invariant when the points $x$ and $y$ are brought into 
coincidence in the final limit, creating a closed link loop. (No%
te that these ``partial derivatives'' do not commute, because ea%
ch actually represents the shifting of an endpoint and the exten%
sion of the path of the link.) Now, acting on the result (%
\protect\ref{eq:Continuum-dUdx}) with the operator \raisebox{0ex%
}[1ex][0ex]{$\protect\displaystyle\partial/\partial{x}{}^{%
\raisebox{-0.25ex}{$\scriptstyle\nu$}}$}\mbox{$\!$}, the product 
rule yields in the first instance \setcounter{Ceqindent}{0}%
\protect\begin{eqnarray}\hspace{-1.3ex}&\displaystyle\mbox{$%
\protect\displaystyle\protect\frac{\partial}{\partial x^\nu}$}%
\mbox{$\protect\displaystyle\protect\frac{\partial}{\partial x^%
\mu}$}^{\:\!}U^{\:\!\!}(y,x,^{\!}P)=-ig\:\!\!\protect\left\{\:\!%
\!\mbox{$\protect\displaystyle\protect\frac{\partial}{\partial x%
^\nu}$}^{\:\!}U^{\:\!\!}(y,x,^{\!}P)\!\protect\right\}\!A_{\mu\:%
\!\!}(x)-ig^{\,}U^{\:\!\!}(y,x,^{\!}P)^{\:\!}\mbox{$\protect
\displaystyle\protect\frac{\partial}{\partial x^\nu}$}^{\:\!}A_{%
\mu\:\!\!}(x).\protect\nonumber\setlength{\Cscr}{\value{CEht}%
\Ctenthex}\addtolength{\Cscr}{-1.0ex}\protect\raisebox{0ex}[%
\value{CEht}\Ctenthex][\Cscr]{}\protect\end{eqnarray}\setcounter
{CEht}{10}We can now substitute (\protect\ref{eq:Continuum-dUdx}%
) itself in for the expression in braces, to obtain \setcounter{%
Ceqindent}{0}\protect\begin{eqnarray}\protect\left.\protect\begin
{array}{rcl}\protect\displaystyle\hspace{-1.3ex}&\protect
\displaystyle\mbox{$\protect\displaystyle\protect\frac{\partial}%
{\partial x^\nu}$}\mbox{$\protect\displaystyle\protect\frac{%
\partial}{\partial x^\mu}$}^{\:\!}U^{\:\!\!}(y,x,^{\!}P)=-ig^{\,%
}U^{\:\!\!}(y,x,^{\!}P)\setcounter{Cbscurr}{20}\setlength{\Cscr}%
{\value{Cbscurr}\Ctenthex}\addtolength{\Cscr}{-1.0ex}\protect
\raisebox{0ex}[\value{Cbscurr}\Ctenthex][\Cscr]{}\hspace{-0ex}{%
\protect\left\{\setlength{\Cscr}{\value{Cbscurr}\Ctenthex}%
\addtolength{\Cscr}{-1.0ex}\protect\raisebox{0ex}[\value{Cbscurr%
}\Ctenthex][\Cscr]{}\protect\right.}\hspace{-0.25ex}\setlength{%
\Cscr}{\value{Cbscurr}\Ctenthex}\addtolength{\Cscr}{-1.0ex}%
\protect\raisebox{0ex}[\value{Cbscurr}\Ctenthex][\Cscr]{}%
\setcounter{CbsD}{\value{CbsC}}\setcounter{CbsC}{\value{CbsB}}%
\setcounter{CbsB}{\value{CbsA}}\setcounter{CbsA}{\value{Cbscurr}%
}\partial_\nu A_{\mu\:\!\!}(x)-igA_{\nu\:\!\!}(x)A_{\mu\:\!\!}(x%
)^{\:\!\!}\setlength{\Cscr}{\value{CbsA}\Ctenthex}\addtolength{%
\Cscr}{-1.0ex}\protect\raisebox{0ex}[\value{CbsA}\Ctenthex][\Cscr
]{}\hspace{-0.25ex}{\protect\left.\setlength{\Cscr}{\value{CbsA}%
\Ctenthex}\addtolength{\Cscr}{-1.0ex}\protect\raisebox{0ex}[%
\value{CbsA}\Ctenthex][\Cscr]{}\protect\right\}}\hspace{-0ex}%
\setlength{\Cscr}{\value{CbsA}\Ctenthex}\addtolength{\Cscr}{-1.0%
ex}\protect\raisebox{0ex}[\value{CbsA}\Ctenthex][\Cscr]{}%
\setcounter{CbsA}{\value{CbsB}}\setcounter{CbsB}{\value{CbsC}}%
\setcounter{CbsC}{\value{CbsD}}\setcounter{CbsD}{1},\setlength{%
\Cscr}{\value{CEht}\Ctenthex}\addtolength{\Cscr}{-1.0ex}\protect
\raisebox{0ex}[\value{CEht}\Ctenthex][\Cscr]{}\protect\end{array%
}\protect\right.\protect\label{eq:Continuum-d2U-dxmu-dxnu}%
\protect\end{eqnarray}\setcounter{CEht}{10}where we can use the 
shorthand \raisebox{0ex}[1ex][0ex]{$\protect\displaystyle\partial
_\nu\equiv\partial/\partial{x}{}^{\raisebox{-0.25ex}{$%
\scriptstyle\nu$}}$} inside the braces because everything in the%
re is a function of $x$ only; the $y$-dependence of (\protect\ref
{eq:Continuum-d2U-dxmu-dxnu}) is now encapsulated completely in 
the link \raisebox{0ex}[1ex][0ex]{$\protect\displaystyle U^{\:\!%
\!}(y,x,^{\!}P)$}. We thus find that \setcounter{Ceqindent}{0}%
\protect\begin{eqnarray}\hspace{-3ex}\hspace{-1.3ex}&%
\displaystyle\protect\left(\!\mbox{$\protect\displaystyle\protect
\frac{\partial}{\partial x^\mu}$}\mbox{$\protect\displaystyle
\protect\frac{\partial}{\partial x^\nu}$}\!-\!\mbox{$\protect
\displaystyle\protect\frac{\partial}{\partial x^\nu}$}\mbox{$%
\protect\displaystyle\protect\frac{\partial}{\partial x^\mu}$}\!%
^{\:\!\!}\protect\right)^{\!}U^{\:\!\!}(y,x,^{\!}P)=-ig^{\,}U^{%
\:\!\!}(y,x,^{\!}P)\setcounter{Cbscurr}{20}\setlength{\Cscr}{%
\value{Cbscurr}\Ctenthex}\addtolength{\Cscr}{-1.0ex}\protect
\raisebox{0ex}[\value{Cbscurr}\Ctenthex][\Cscr]{}\hspace{-0ex}{%
\protect\left\{\setlength{\Cscr}{\value{Cbscurr}\Ctenthex}%
\addtolength{\Cscr}{-1.0ex}\protect\raisebox{0ex}[\value{Cbscurr%
}\Ctenthex][\Cscr]{}\protect\right.}\hspace{-0.25ex}\setlength{%
\Cscr}{\value{Cbscurr}\Ctenthex}\addtolength{\Cscr}{-1.0ex}%
\protect\raisebox{0ex}[\value{Cbscurr}\Ctenthex][\Cscr]{}%
\setcounter{CbsD}{\value{CbsC}}\setcounter{CbsC}{\value{CbsB}}%
\setcounter{CbsB}{\value{CbsA}}\setcounter{CbsA}{\value{Cbscurr}%
}\partial_\mu A_{\nu\:\!\!}(x)-\partial_\nu A_{\mu\:\!\!}(x)-ig[%
A_{\mu\:\!\!}(x),A_{\nu\:\!\!}(x)]\setlength{\Cscr}{\value{CbsA}%
\Ctenthex}\addtolength{\Cscr}{-1.0ex}\protect\raisebox{0ex}[%
\value{CbsA}\Ctenthex][\Cscr]{}\hspace{-0.25ex}{\protect\left.%
\setlength{\Cscr}{\value{CbsA}\Ctenthex}\addtolength{\Cscr}{-1.0%
ex}\protect\raisebox{0ex}[\value{CbsA}\Ctenthex][\Cscr]{}\protect
\right\}}\hspace{-0ex}\setlength{\Cscr}{\value{CbsA}\Ctenthex}%
\addtolength{\Cscr}{-1.0ex}\protect\raisebox{0ex}[\value{CbsA}%
\Ctenthex][\Cscr]{}\setcounter{CbsA}{\value{CbsB}}\setcounter{Cb%
sB}{\value{CbsC}}\setcounter{CbsC}{\value{CbsD}}\setcounter{CbsD%
}{1}.\hspace{-3ex}\protect\nonumber\setlength{\Cscr}{\value{CEht%
}\Ctenthex}\addtolength{\Cscr}{-1.0ex}\protect\raisebox{0ex}[%
\value{CEht}\Ctenthex][\Cscr]{}\protect\end{eqnarray}\setcounter
{CEht}{10}We now note that the expression in braces is just the 
definition of \raisebox{0ex}[1ex][0ex]{$\protect\displaystyle F_%
{\:\!\!\mu\nu}(x)$}: \setcounter{Ceqindent}{0}\protect\begin{eqn%
array}\hspace{-1.3ex}&\displaystyle F_{\:\!\!\mu\nu}(x)\equiv
\partial_\mu A_{\nu\:\!\!}(x)-\partial_\nu A_{\mu\:\!\!}(x)-ig[A%
_{\mu\:\!\!}(x),A_{\nu\:\!\!}(x)],\protect\nonumber\setlength{%
\Cscr}{\value{CEht}\Ctenthex}\addtolength{\Cscr}{-1.0ex}\protect
\raisebox{0ex}[\value{CEht}\Ctenthex][\Cscr]{}\protect\end{eqnar%
ray}\setcounter{CEht}{10}so that \setcounter{Ceqindent}{0}%
\protect\begin{eqnarray}\protect\left.\protect\begin{array}{rcl}%
\protect\displaystyle\hspace{-1.3ex}&\protect\displaystyle
\protect\left(\!\mbox{$\protect\displaystyle\protect\frac{%
\partial}{\partial x^\mu}$}\mbox{$\protect\displaystyle\protect
\frac{\partial}{\partial x^\nu}$}\!-\!\mbox{$\protect
\displaystyle\protect\frac{\partial}{\partial x^\nu}$}\mbox{$%
\protect\displaystyle\protect\frac{\partial}{\partial x^\mu}$}\!%
^{\:\!\!}\protect\right)^{\!}U^{\:\!\!}(y,x,^{\!}P)=-ig^{\,}U^{%
\:\!\!}(y,x,^{\!}P)^{\:\!}F_{\:\!\!\mu\nu}(x).\setlength{\Cscr}{%
\value{CEht}\Ctenthex}\addtolength{\Cscr}{-1.0ex}\protect
\raisebox{0ex}[\value{CEht}\Ctenthex][\Cscr]{}\protect\end{array%
}\protect\right.\protect\label{eq:Continuum-dmu-dnu-U}\protect
\end{eqnarray}\setcounter{CEht}{10}The $y$-derivatives in (%
\protect\ref{eq:Continuum-DefineCGx}) similarly spit out a facto%
r of \raisebox{0ex}[1ex][0ex]{$\protect\displaystyle F_{\:\!\!\mu
\nu}(y)$}, so that, when we take the limit \raisebox{0ex}[1ex][0%
ex]{$\protect\displaystyle y\rightarrow x$} with \raisebox{0ex}[%
1ex][0ex]{$\protect\displaystyle P^{\:\!\!}\rightarrow0$}, we ob%
tain \setcounter{Ceqindent}{0}\protect\begin{eqnarray}\protect
\left.\protect\begin{array}{rcl}\protect\displaystyle\hspace{-1.%
3ex}&\protect\displaystyle C_{\:\!\!G}(x)\equiv-{g}{}^{\raisebox
{-0.25ex}{$\scriptstyle2\,$}}\mbox{Tr$\:$}^{\!}F_{\:\!\!\mu\nu}(%
x)^{\:\!}F^{\mu\nu\:\!\!}(x).\setlength{\Cscr}{\value{CEht}%
\Ctenthex}\addtolength{\Cscr}{-1.0ex}\protect\raisebox{0ex}[%
\value{CEht}\Ctenthex][\Cscr]{}\protect\end{array}\protect\right
.\protect\label{eq:Continuum-CGxFsquared}\protect\end{eqnarray}%
\setcounter{CEht}{10}In other words, the pure gauge part of the 
action is simply proportional to \raisebox{0ex}[1ex][0ex]{$%
\protect\displaystyle C_{\:\!\!G}(x)$} of (\protect\ref{eq:Conti%
nuum-DefineCGx}), which is constructed from an infinitesimal lin%
k, by shifting each endpoint in both the $\mu$ and $\nu$ directi%
ons, forming the appropriate commutators of these ``derivatives%
''\mbox{$\!$}, and then bringing the two ends of the infinitesim%
al link together into a gauge-invariant loop.\par\refstepcounter
{section}\vspace{1.5\baselineskip}\par{\centering\bf\thesection. 
Implementing the gauge action optimally on the lattice\\*[0.5%
\baselineskip]}\protect\indent\label{sect:Lattice}In the previou%
s section we found that the Lagrangian of a continuum gauge theo%
ry can be written completely in terms of the fermion fields and 
``continuum links'' (Schwinger line integrals), without need to 
explicitly use the conventional gauge field $A_\mu$ at all. When 
we sample this formalism onto the lattice, we continue to have u%
se of the discrete links \setcounter{Ceqindent}{0}\protect\begin
{eqnarray}\protect\left.\protect\begin{array}{rcl}\protect
\displaystyle\hspace{-1.3ex}&\protect\displaystyle U_{\:\!\!\mu}%
(x)\equiv U^{\:\!\!}(x\!+\!a\hat\mu,x,P_{x,\,x+a\hat\mu}),%
\setlength{\Cscr}{\value{CEht}\Ctenthex}\addtolength{\Cscr}{-1.0%
ex}\protect\raisebox{0ex}[\value{CEht}\Ctenthex][\Cscr]{}\protect
\end{array}\protect\right.\protect\label{eq:Lattice-GaugeLinks}%
\protect\end{eqnarray}\setcounter{CEht}{10}where \raisebox{0ex}[%
1ex][0ex]{$\protect\displaystyle P_{x,\,x+a\hat\mu}$} is a strai%
ght-line path from the site at position $x$ to the adjacent site 
in the $\mu$-direction at position \raisebox{0ex}[1ex][0ex]{$%
\protect\displaystyle x+a\hat\mu$}.\par For the fermionic $D$-sl%
ash part of the action, the expression (\protect\ref{eq:Continuu%
m-DefineCFx}) tells us to join $\bar\psi(y)\gamma^\mu$ to $\psi(%
x)$ with a link, and then to take the ``derivative'' with respec%
t to \raisebox{0ex}[1ex][0ex]{$\protect\displaystyle x^{\mu\!}$}%
, after which we can put $y$ back at the position $x$. A moment'%
s thought shows that this is precisely what we already do on the 
lattice, when we use the naive\ derivative or Wilson fermions: w%
hen we take the discrete derivative of $\psi(x)$, we join it bac%
k to the $\bar\psi(x)\gamma^\mu$ with the appropriate link matri%
ces. Using the SLAC derivative simply requires that each term of 
$\partial_\mu\psi$ be connected back to $\bar\psi(x)\gamma^\mu$ 
with the appropriate link---which will, for distances greater th%
an one lattice unit, consist of products of the elementary links 
(\protect\ref{eq:Lattice-GaugeLinks}). A stochastic implementati%
on of this operator will not threaten gauge invariance, because 
each term is separately connected back to $\bar\psi(x)\gamma^\mu
$ with the appropriate link. This is, in fact, precisely how the 
operator has now been implemented in the MILC code. (To ensure m%
aximal performance, the ``integrals'' of the links along each co%
ordinate direction are first computed, so that the extraction of 
these ``long links'' reduces to simply a reading of these ``inte%
grated links'' at the endpoints, sometimes supplemented by the u%
se of a ``Wilson line'' if the link goes ``around the back'' of 
the lattice.)\par The same thing can be done with the expression 
(\protect\ref{eq:Continuum-DefineCGx}) for the pure gauge part o%
f the action. If we expand out the parentheses we obtain four te%
rms, corresponding to the derivatives at the $x$ and $y$ ends be%
ing taken in opposite orders, and the path-reversals of these tw%
o combinations. As with the fermionic action (\protect\ref{eq:Co%
ntinuum-DefineCFx}), we are free to apply whichever definition o%
f the derivative operator we desire.\par Even the naive\ derivat%
ive operator provides a richer variety of contributions than wha%
t one might expect. Clearly, some of the resulting diagrams will 
correspond to the usual plaquette contributions. Others will enc%
lose no area whatsoever (the shifted endpoints will retrace each 
other), and thus simply contribute to the constant term which th%
e sum of plaquettes is subtracted from. Over and above these low%
est-order contributions, however, we also obtain rectangles cove%
ring two lattice rectangles, as well as ``bow tie'' and ``collar%
'' diagrams, in which we somehow have to decide how to connect t%
he two ends that are displaced diagonally from each other.\par T%
he simplest ansatz to apply in this case---and the only one that 
is computationally feasible for higher-order diagrams---is that 
an unspecified link is split into two equal contributions, one w%
hich tracks from $x$ to $y$ along the $\mu$-direction and then t%
he $\nu$-direction, and the other which tracks from $x$ to $y$ a%
long the $\nu$-direction and then the $\mu$-direction. If we mak%
e this ansatz, then no diagram requires the use of more than six 
(long) lattice links, regardless of whether we use the naive\ de%
rivative or the optimal SLAC derivative. This not only makes the 
calculation of each diagram computationally simple (provided tha%
t we have available the ``integrated long link'' library describ%
ed above, as has already been implemented in the MILC code), but 
moreover allows the diagrams to be generated, sorted, simplified%
\ and stored efficiently.\par This latter consideration is of im%
portance to us if we wish to use the SLAC derivative, in order t%
o eliminate scaling artifacts, because the expression (\protect
\ref{eq:Continuum-DefineCGx}) requires the use of four derivativ%
e operators, which implies that the number of possible diagrams 
is of the same order of magnitude as the total number of sites o%
n the lattice. Clearly, a stochastic implementation of the opera%
tor is called for! Fortunately, the coefficients for the bulk of 
these diagrams are as small as the diagrams themselves are numer%
ous, precisely because the coefficient of each term in the SLAC 
derivative operator falls off inversely with distance. In practi%
ce, it may prove worthwhile to truncate the number of stored dia%
grams, so that those with a coefficient smaller than some thresh%
old are simply discarded, because they contribute so negligibly 
on the average, compared to the other sources of noise in the Mo%
nte Carlo process. However, this is a practical trade-off that c%
an be experimented with when the operator has been implemented; 
there is no algorithmic or computational obstacle to computing a%
nd storing all diagrams, provided that the list is sorted in dec%
reasing order of probability (as is currently done with the stoc%
hastic $D$-slash and antialiasing operators), because in the vas%
t majority of cases the vast bulk of this list will never be tra%
versed.\par\refstepcounter{section}\vspace{1.5\baselineskip}\par
{\centering\bf\thesection. Conclusions\\*[0.5\baselineskip]}%
\protect\indent\label{sect:Conclusions}I have argued in this pap%
er that it is possible to construct the third and final plank of 
a new paradigm for lattice computations, that holds out the prom%
ise of yielding results that do not possess any scaling artifact%
s over and above those physically created by putting gauge theor%
y onto a finite lattice.\par This third plank will be implemente%
d shortly using the MILC code, and put together with the optimal 
$D$-slash and antialiasing filters. Physics results are expected 
by mid-2005.\par\vspace{1.5\baselineskip}\par{\centering\bf Ackn%
owledgments\\*[0.5\baselineskip]}\protect\indent This work was i%
n part based on the MILC collaboration's public lattice gauge th%
eory code. See [\ref{cit:MILC}]. Helpful discussions with A.~A.~%
Rawlinson, B.~H.~J.~McKellar, C.~Davies, H.~Neuberger\ and D.~Ko%
ks are gratefully acknowledged.\par\vspace{1.5\baselineskip}\par
{\centering\bf References\\*[0.5\baselineskip]}{\protect\mbox{}}%
\vspace{-\baselineskip}\vspace{-2ex}\settowidth\CGDnum{[\ref{cit%
last}]}\setlength{\CGDtext}{\textwidth}\addtolength{\CGDtext}{-%
\CGDnum}\begin{list}{Error!}{\setlength{\labelwidth}{\CGDnum}%
\setlength{\labelsep}{0.75ex}\setlength{\leftmargin}{\labelwidth
}\addtolength{\leftmargin}{\labelsep}\setlength{\rightmargin}{0e%
x}\setlength{\itemsep}{0ex}\setlength{\parsep}{0ex}}\protect
\frenchspacing\setcounter{CBtnc}{1}\addtocounter{CBcit}{1}\item[%
\hfill{[}\arabic{CBcit}{]}]\renewcommand\theCscr{\arabic{CBcit}}%
\protect\refstepcounter{Cscr}\protect\label{cit:Press1988}W.~H.~%
Press, B.~P.~Flannery, S.~A.~Teukolsky\ and W.~T.~Vetterling, 
\renewcommand\theCscr{Press, Flannery, Teukolsky\ and Vetterling%
}\protect\refstepcounter{Cscr}\protect\label{au:Press1988}%
\renewcommand\theCscr{1988}\protect\refstepcounter{Cscr}\protect
\label{yr:Press1988}\protect\emph{Numerical Recipes in C} (Cambr%
idge University Press, Cambridge, 1988), \S12.1.\addtocounter{CB%
cit}{1}\item[\hfill{[}\arabic{CBcit}{]}]\renewcommand\theCscr{%
\arabic{CBcit}}\protect\refstepcounter{Cscr}\protect\label{cit:C%
ostella2004}J.~P.~Costella, \renewcommand\theCscr{Costella}%
\protect\refstepcounter{Cscr}\protect\label{au:Costella2004}%
\renewcommand\theCscr{2004}\protect\refstepcounter{Cscr}\protect
\label{yr:Costella2004}{}hep-lat/0404019.\addtocounter{CBcit}{1}%
\item[\hfill{[}\arabic{CBcit}{]}]\renewcommand\theCscr{\arabic{C%
Bcit}}\protect\refstepcounter{Cscr}\protect\label{cit:MILC}MILC 
collaboration, \renewcommand\theCscr{MILC collaboration}\protect
\refstepcounter{Cscr}\protect\label{au:MILC}\renewcommand\theCscr
{2004}\protect\refstepcounter{Cscr}\protect\label{yr:MILC}{}http%
:/\raisebox{0ex}[1ex][0ex]{$\protect\displaystyle\!$}/www.physic%
s.utah.edu/\raisebox{0ex}[1ex][0ex]{$\protect\displaystyle\!\sim
$}detar/milc/.\addtocounter{CBcit}{1}\item[\hfill{[}\arabic{CBci%
t}{]}]\renewcommand\theCscr{\arabic{CBcit}}\protect
\refstepcounter{Cscr}\protect\label{cit:Costella2002a}J.~P.~Cost%
ella, \renewcommand\theCscr{Costella}\protect\refstepcounter{Csc%
r}\protect\label{au:Costella2002a}\renewcommand\theCscr{2002a}%
\protect\refstepcounter{Cscr}\protect\label{yr:Costella2002a}{}h%
ep-lat/0207008.\addtocounter{CBcit}{1}\item[\hfill{[}\arabic{CBc%
it}{]}]\renewcommand\theCscr{\arabic{CBcit}}\protect
\refstepcounter{Cscr}\protect\label{cit:Drell1976}S.~D.~Drell, M%
.~Weinstein\ and S.~Yankielowicz, \renewcommand\theCscr{Drell, W%
einstein\ and Yankielowicz}\protect\refstepcounter{Cscr}\protect
\label{au:Drell1976}\renewcommand\theCscr{1976}\protect
\refstepcounter{Cscr}\protect\label{yr:Drell1976}\protect\emph{P%
hys. Rev.~D}\ {\bf14}\ (1976) 487; 1627.\addtocounter{CBcit}{1}%
\item[\hfill{[}\arabic{CBcit}{]}]\renewcommand\theCscr{\arabic{C%
Bcit}}\protect\refstepcounter{Cscr}\protect\label{cit:Costella20%
02b}J.~P.~Costella, \renewcommand\theCscr{Costella}\protect
\refstepcounter{Cscr}\protect\label{au:Costella2002b}%
\renewcommand\theCscr{2002b}\protect\refstepcounter{Cscr}\protect
\label{yr:Costella2002b}{}hep-lat/0207015.\renewcommand\theCscr{%
\arabic{CBcit}}\protect\refstepcounter{Cscr}\protect\label{citla%
st}\settowidth\Cscr{~[\ref{cit:Costella2002b}]}\end{list}\par\end
{document}